\documentclass[12pt]{iopart}
\usepackage{graphicx,cite}

\begin{document} 
\bibliographystyle{jpb}

\title{Sound wave propagation in strongly elongated fermion clouds 
at finite collisionality}
\author{P Capuzzi, P Vignolo, F  Federici and M  P Tosi}
\address{NEST-INFM-CNR and
Classe di Scienze, Scuola Normale Superiore, I-56126 Pisa, Italy}
\eads{\mailto{capuzzi@sns.it}, \mailto{vignolo@sns.it},
\mailto{fr.federici@sns.it}, \mailto{tosim@sns.it}}

\begin{abstract}
We evaluate the transition from zero-sound to first-sound behaviour
with increasing collisionality in the propagation of density waves
through an ultracold gaseous mixture of fermionic atoms confined
in the normal state inside a cigar-shaped harmonic trap.
We study for this purpose the evolution of
the one-body distribution functions associated with a density
perturbation generated in the central region of the cloud,
as obtained by
solving numerically the Vlasov-Landau equations.
We examine a variety of trap anisotropies and of repulsive or attractive
interaction strengths between the components of the mixture,
and the speed of propagation of the density disturbance is
found to decrease in both cases as the magnitude of the coupling strength
is increased. The results are compared with the values of the speed
of zero sound and of first sound, as obtained analytically from the
limit of vanishing collisionality and from linearized hydrodynamics.
The main effects of the quasi-one-dimensional confinement are 
the stabilization of zero-sound excitations in the attractive regime before collapse
and the lowering of the hydrodynamic sound velocity by a factor
$\sqrt{3/5}$ relative to three-dimensional behaviour.

\end{abstract}

\submitto{\JPB}
\pacs{03.75.Ss, 05.30.Fk, 02.70.Ns}

\maketitle

\section{Introduction}
Several experiments on mixed Fermi gases consisting of two components
with different pseudospins have probed their collisional properties
both in the normal \cite{Gensemer2001a} and in the superfluid state
\cite{Kinast2004a,Bartenstein2004a}.  In the experiments performed by
the JILA group, the collisionality of a $^{40}$K mixture in the normal
state was tuned by varying either the atomic density
\cite{Gensemer2001a,DeMarco2002a} or the off-resonant value of the
scattering length \cite{Loftus2002a,Regal2003b}. Gensemer and Jin
\cite{Gensemer2001a} examined  the frequency and damping of the
dipolar modes of the mixture as functions of its  collisionality,
showing  that, while in the collisionless limit the two components
oscillate independently with vanishing damping, on  increasing the
strength of the mutual repulsive interactions the damping first
increases and then decreases until in the hydrodynamic limit the two
components oscillate together at the same frequency without damping.
The same behaviour has been obtained from  numerical simulations
\cite{Toschi2003a,Toschi2004a}.  The absence of damping in the two
opposite limits signals the existence of long-lived collective
sound-wave excitations in suitable system geometries, the so-called
zero sound in the collisionless limit and the hydrodynamic first sound
in the strongly collisional regime.

More recently attention has focused on sound-wave propagation in a
Fermi gas with attractive interactions  ranging in strength from the
BCS to the BEC regime
\cite{Ho2004a,Heiselberg2005b,Capuzzi2006a,Urban2005a}.  Below the
critical temperature for the transition to a superfluid state, the
presence of both normal and superfluid components leads to two
acoustic modes corresponding to the propagation of density waves and
thermal waves \cite{Ho2004a,Heiselberg2005a}.  These modes contain
information on the equation of state of the gas and on the geometry of
the system \cite{Capuzzi2006a} and impose constraints on the
appropriate theoretical models to describe the BCS-BEC crossover
\cite{Heiselberg2005b}.  Above the critical temperature the speed of
propagation of a density perturbation still gives information about
the shape of the equilibrium density profiles and about the
collisionality of the gas relative to the frequency of the
perturbation.  

The behaviour that we have recalled above for the attenuation of sound
waves as a function of collisionality was predicted in the early work
of Landau on normal liquid $^3$He \cite{Pines94}. In the  
experiments of Abel {\it et al.}~\cite{Abel1966a} on this liquid, 
the collisionality was explored by tuning the
frequency of an acoustic disturbance.  The experiment demonstrated
that the transition from collisional to collisionless dynamics is
signalled by a drastic change in the frequency dependence of the
attenuation and showed that the ratio of the speeds of zero and first
sound is very much smaller than the value $\sqrt{3}$ expected for a
homogeneous fluid at weak coupling. In a dilute Fermi gas one should
expect  that the interactions introduce instead only slight modifications to the bare
sound velocities in the ideal Fermi gas.

The main purpose of the present work is to provide direct theoretical
evidence for the transition from zero to first sound in a normal
Fermi gas by evaluating the propagation of density
waves in a two-component mixture confined inside an elongated
cigar-shaped harmonic trap. We have in mind an experiment in which the
collisionality could be varied by tuning the scattering length $a$ as can be
realized by exploiting Feshbach resonances~\cite{Loftus2002a} and the
density perturbation could be generated by a laser beam focused at
the centre of the trap \cite{Andrews1997a}.   We
perform numerical simulations  of the dynamics of the gas, in which collisions
are taken into account by solving the Vlasov-Landau equations
(VLE) for the one-body distribution functions of the two fermionic 
components.  We observe that on increasing the scattering
cross-section the velocity of propagation of
the density distortions, which is given by the Fermi velocity $v_{\rm
F}$ in the collisionless limit, diminishes for both attractive and
repulsive coupling. If the mean-field corrections are negligible, the
hydrodynamic limit for the confined gas in the elongated trap yields a
propagation velocity 
$v_{\mathrm{F}}/\sqrt{5}$ instead of $v_{\mathrm{F}}/\sqrt{3}$ as
appropriate to a homogeneous gas. Most remarkably, the zero-sound
excitation is found to be stable even at weak attractive couplings,
before the attractions drive the gas to collapse. 

The paper is organized as follows. In Sec.\ \ref{sec:system} we
introduce the fermion mixture under study and discuss the 
dynamical regimes that can be realized in a dilute gaseous state, a
detailed analytic calculation for zero sound in a cylindrical trap
being reported in an Appendix.  In Sec.\
\ref{sec:propagation} we present our treatment of a mixture
with finite collisionality and discuss our numerical results for the
velocity of propagation of density waves. Finally, Sec.\
\ref{sec:summary} offers a summary and the main conclusions of the work.

\section{\label{sec:system}The model system} 

We consider a mixture of two species of fermionic atoms of mass $m$
confined in a harmonic trap of the form 
\begin{equation}
V_{\rm ext}(\bi{r}) = \frac{1}{2}\,m\,\omega_{\perp}^2
(r_{\perp}^2+\lambda^2\,z^2)\,,
\label{Eq:pote}
\end{equation}
where $\lambda=\omega_z/\omega_{\perp}$ is the anisotropy
parameter.  The atoms of each species are spin-polarized and the gas
is highly dilute,
so that at very low temperatures only interactions between atoms of
different species are allowed. 
The gas is described in terms of the one-body
distribution functions $f^{(s)}(\bi{r},\bi{p},t)$ 
for each species $s$ and the particle densities are obtained by
performing an integration over momentum space,
\begin{equation}
n^{(s)}(\bi{r},t) = \int \frac{d^3\bi{p}}{(2\pi\hbar)^3} 
f^{(s)}(\bi{r},\bi{p},t). 
\label{eq:nsigma}
\end{equation}
The distribution functions at equilibrium  
are given by the local Fermi-Dirac distributions
\begin{equation}
f_0^{(s)} (\bi{r},\bi{p}) = \left\{\exp\left[\beta \left(
\frac{p^2}{2m}+U^{(s)}(\bi{r}) - \mu^{(s)} \right) \right]
 + 1\right\}^{-1}, 
\label{eq:fequilibrio} 
\end{equation}  
where $\beta=1/k_BT$ with $T$ the  temperature, 
$\mu^{(s)}$ is the chemical potential ensuring the
normalization $N_{s}=\int f^{(s)}(\bi{r},
\bi{p})\,d^3r\,d^3p/h^3 $, and the mean-field potentials read 
\begin{equation}
U^{(s)}(\bi{r}) =
V_{\rm{ext}}(\bi{r}) + g\,n^{(\bar{s})}(\bi{r}).
\label{eq:Ueff}
\end{equation}
Here ${\bar{s}}$ denotes the component different from $s$
and $g=4\pi\hbar^2\,a/m$ is the coupling strength between atoms of
different species, with $a$ being their $s$-wave scattering length. 

The calculation of the equilibrium density profiles $n_0^{(s)}$ 
requires the self-consistent solution of equations
(\ref{eq:nsigma})-(\ref{eq:Ueff}) for each set of system parameters.  
At $T=0$ the results  of the
Thomas-Fermi approximation are recovered, giving $n_0^{(s)}$ as the solution of the equation 
$\mu(n_0^{(s)}) + V_{\rm{ext}}(\bi{r})= \mu^{(s)}$, with 
$\mu(n)$ the equation of state of the gas
\begin{equation}
\mu(n_0^{(s)}) = \frac{\hbar^2}{2m}\left(6\pi^2\,n_0^{(s)}\right)^{2/3} +
g\,n_0^{(\bar{s})} .
\label{eq:eos}
\end{equation}
We shall study a mixture of 2000 $^{40}$K atoms at
$T=0.2\,T_F$, $T_F$ being the Fermi temperature. The atoms are 
equally distributed  in two spin states and confined inside a trap with
radial frequency $\omega_{\perp}=100\,$s$^{-1}$. Hereafter
we will remove the indices $s$ and $\bar{s}$ when there is no
ambiguity  and set $\mu^{(s)}=\mu^{(\bar{s})}=\mu$ and $n^{(s)}=n^{(\bar{s})}=n$.

\subsection{Sound propagation}
\label{sec:sound_propa}
The time evolution of the one-body distributions strongly depends  on
the frequency of the inter-species collisions. When this is much lower than 
the trap frequency, the evolution of the gas is governed by deformations of
the Fermi sphere and the dynamics is termed collisionless. 
This is the regime of zero sound. In a homogeneous mixture 
the collisionless sound velocity is given by $c_0= \eta v_{\mathrm{F}}$, 
with the Fermi velocity $v_{\mathrm{F}}=[2(\mu-g\,n_0)/m]^{1/2}$.
The ratio $\eta$ can be calculated in Landau's Fermi-liquid theory describing the
mean-field interactions through appropriate Landau parameters.
For a two-component Fermi gas with negligible interactions between
atoms of the same spin, $\eta$ satisfies the relation \cite{Pines94,Akdeniz2003b}
\begin{equation}
\frac{\eta}{2}\ln\frac{\eta+1}{\eta-1}-1=
\frac{\pi\hbar}{2mv_{\mathrm{F}}\,a}\,.
\label{eq:eta3D}
\end{equation}
In a trap one may use a local-density approximation and
assign the position-dependent Fermi velocity $v_{\mathrm{F}}(\bi{r}) =
[2(\mu-V_{\rm{ext}}({\bi{r}})-gn_0 (\bi{r}))/m]^{1/2}$ to the
atoms, provided the variation of the density profile is
sufficiently smooth. In this way the confinement is accounted for only
through the position dependence of the density profiles.  

However, a very tight confinement may drastically modify the density
of states of the gas and thus the dynamic of density perturbations.
The zero sound velocity $\tilde c_0$ in strongly elongated traps
($\lambda\rightarrow 0$ and $V_{\rm{ext}}({\bi{r}}) \rightarrow
m\omega_\perp^2r_\perp^2/2$), in a regime where the degrees of freedom
in the azimuthal plane are frozen by the confinement, is given by (see
Appendix)
\begin{equation}
\tilde{c}_0=v_{\mathrm{F}}^0\left(1+\frac{2 \hbar\, a}{\pi \tilde a_\perp^2 
m v_{\mathrm{F}}^0}\right)^{1/2},
\label{eq:zerosound}
\end{equation}
with $v_{\mathrm{F}}^0$ being the local Fermi velocity at the centre 
of the trap and $2\pi\tilde a_\perp^2$ the radial section of the atomic cloud.
This velocity coincides with that of a strictly one-dimensional (1D) 
Fermi gas \cite{Apostol1992a,Hernandez2002a} with an effective 
coupling $2\hbar^2a/(m\tilde{a}_{\perp}^2)$ embodying the
transverse confinement \cite{Salasnich2002a}.  

Interactions between the two components make the sound velocity
deviate from $v_{\mathrm{F}}$ by only a few percent for couplings
up to  $a=2\times10^4$ Bohr radii for both the homogeneous
and the cigar-shaped gas \cite{Akdeniz2003b}. 
The effects of the
interactions  have also been investigated in the random-phase
approximation for a Fermi gas in a spherical trap in the
collisionless regime \cite{Capuzzi2001a,Bruun2001b}.  The main outcome is
the appearance of a fragmented excitation spectrum at strong couplings and of
Landau-damped modes for attractive interactions. However, the
confinement introduces a discrete single-particle excitation spectrum
that may weaken the strength of the damping. In a cigar-shaped
confinement, the sound velocity  given in  equation (\ref{eq:zerosound}) is
well defined for repulsive interactions as well as sufficiently weak attractive
interactions, analogously to zero sound in a 1D
Fermi gas \cite{Hernandez2002a}.

When the dynamics is instead governed by the hydrodynamic equations, 
the (first)
sound velocity in a homogeneous gas
is given by 
\begin{equation}
c_1= \frac{v_{\mathrm{F}}}{\sqrt{3}}\,\left(1+ \frac{2 a}{\pi} \frac{m
v_{\mathrm{F}}}{\hbar}\right)^{1/2}, 
\label{eq:cs}
\end{equation}
showing that the explicit dependence on the scattering length is qualitatively
similar to that in equation (\ref{eq:zerosound}). 
In addition to the density modes, an
oscillation of the  spin density $\xi=n^{(s)}-n^{(\bar{s})}$ may develop
 depending on how the
gas is perturbed \cite{Vichi1999a,Bruun1999b,Amoruso2000a}.  
In fact, for strong repulsions the two species tend to
segregate into two non-overlapping regions.
For attractive interactions, the system will become unstable and experience a
sudden collapse when the sound
velocity vanishes.   However, at very
low temperatures the collapse may be prevented by a transition to a
superfluid phase \cite{Kinast2004a,Chin2004a,Zwierlein2005a}
or by pairing of the atoms into molecular dimers
\cite{Greiner2003a,Jochim2003b,Zwierlein2003a}.

With the purpose of examining the confinement effects on the
propagation of first sound, we consider again a radially confined
Fermi gas and  assume a density wave travelling in the axial $z$
direction. In this case we have shown \cite{Capuzzi2006a} that the
first sound velocity $\tilde{c}_1$ depends on the equation of state
through the relation 
\begin{equation}
\tilde{c}_1 = \left(\frac{1}{m}\,\int\,n_0\,d^2r_{\perp} \Bigr/ \int
(\partial \mu/\partial
n|_{n_0})^{-1}\,d^2r_{\perp}\right)^{1/2}
\label{eq:cstrap}
\end{equation}
where the integration is on the azimuthal plane.  Equation
(\ref{eq:cstrap}) reduces to equation (\ref{eq:cs}) for a homogeneous
Fermi gas and yields $\tilde{c}_1 = v_{\mathrm{F}}^0/\sqrt{5}$ for the
cigar-shaped Fermi gas without mean-field interactions. More generally,
numerical integration can be carried out once the equilibrium density
is known. 

\section{\label{sec:propagation}Density waves  at finite collisionality}
Neither the collisionless nor the hydrodynamic
approach is appropriate when the number of collisions is sizable but 
not sufficient to establish
local equilibrium in the gas. One must then resort to kinetic equations
and can use a
Boltzmann equation method to take into account binary collisions
in a semiclassical manner. The evolution of the
one-body distributions is determined by the VLE 
\begin{equation}
\partial_t f^{(s)} + \frac{\bi{p}}{m}\cdot\nabla_{\bi{r}}
f^{(s)}-\nabla_{\bi{r}}U^{(s)}\cdot\nabla_{\bi{p}}f^{(s)} 
= C^{(s)}[f^{(s)}, f^{(\bar{s})}],
\label{eq:vlasov}
\end{equation}
where the collision integral $C^{(s)}$ reads
\begin{eqnarray}
\fl C^{(s)} = \frac{\sigma}{4\pi(2\pi\hbar)^3}\int\!\! d^3p_2\,d\Omega_f
v\,[(1-f^{(s)})(1-f_2^{(\bar{s})})f_3^{(s)}
f_4^{(\bar{s})} \nonumber \\ 
 - f^{(s)}f^{(\bar{s})}_2(1-f_3^{(s)})
(1-f_4^{(\bar{s})})].
\label{eq:integralcoll}
\end{eqnarray}
Here $f^{(s)}\equiv f^{(s)}({\bi{r}},{\bi{p}},t)$ and $f_i^{(s)}\equiv
f^{(s)}({\bi{r}},{\bi{p}}_i,t)$, $d\Omega_f$ is the element of solid
angle for the outgoing relative momentum $\bi{p}_3-\bi{p}_4$,
$\sigma=4\pi\,a^2$ is the scattering cross-section, and 
$v=|\bi{v}-\bi{v}_2|$ is the relative velocity of the incoming
particles. The collision satisfies conservation of momentum
($\bi{p}+\bi{p}_2 = \bi{p}_3 + \bi{p}_4$) and energy
($\varepsilon + \varepsilon_2 = \varepsilon_3 + \varepsilon_4$), with
$\varepsilon_{i}= p_{i}^2/2m + U^{(s_i)}$.

\begin{figure}[tb]
\begin{tabular}{ccc}
\includegraphics[width=0.32\linewidth,clip=true]{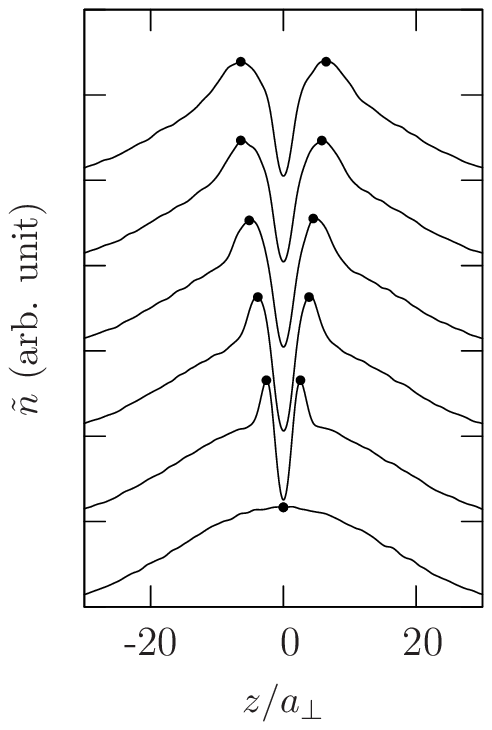}&
\includegraphics[width=0.32\linewidth,clip=true]{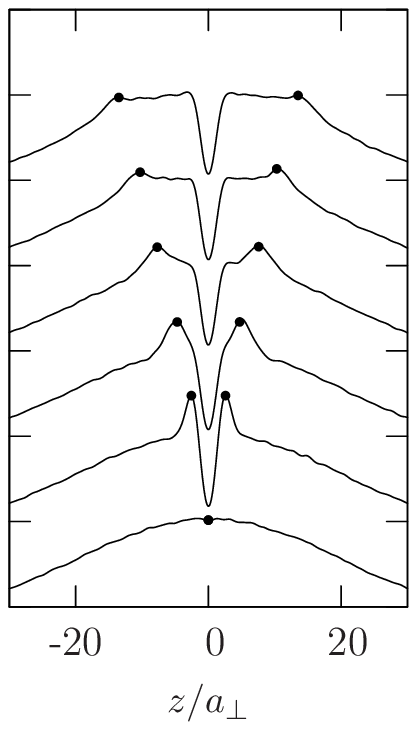}&
\includegraphics[width=0.32\linewidth,clip=true]{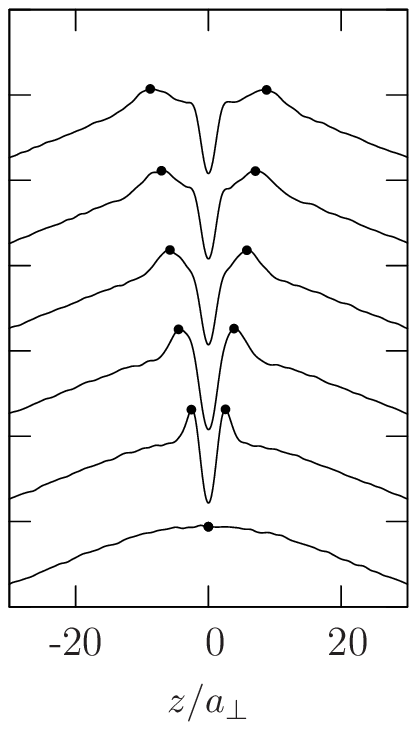}
\end{tabular}
\caption{\label{fig:prof_shifted}Time evolution of the integrated
density profile $\tilde{n}(z)$ as a function of axial position
$z$ (in units of
$a_{\perp}=\sqrt{\hbar/(m\omega_{\perp})}$) for 2000 $^{40}$K atoms in
a trap with $\lambda=0.1$. The left, middle, and right panels
correspond to $a=-1.5\times 10^4, 80$, and $2\times 10^4$ Bohr radii,
respectively.  The bottom profile refers to the initial equilibrium
state and the
subsequent ones are separated by time intervals of
$0.5/\omega_{\perp}=5\,{\rm ms}$.  The
later profiles have been vertically displaced for the sake of visibility.}
\end{figure}

\begin{figure}
\begin{center}
\includegraphics[width=0.7\columnwidth,clip=true]{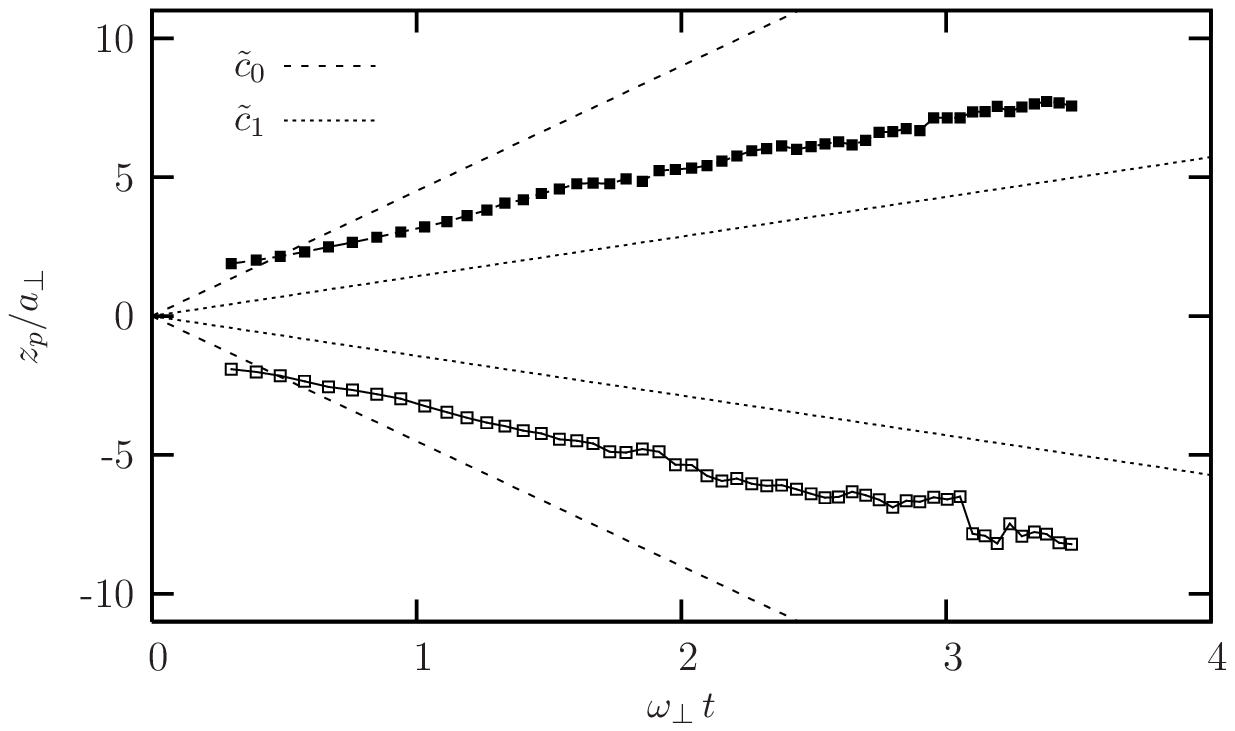}
\includegraphics[width=0.7\columnwidth,clip=true]{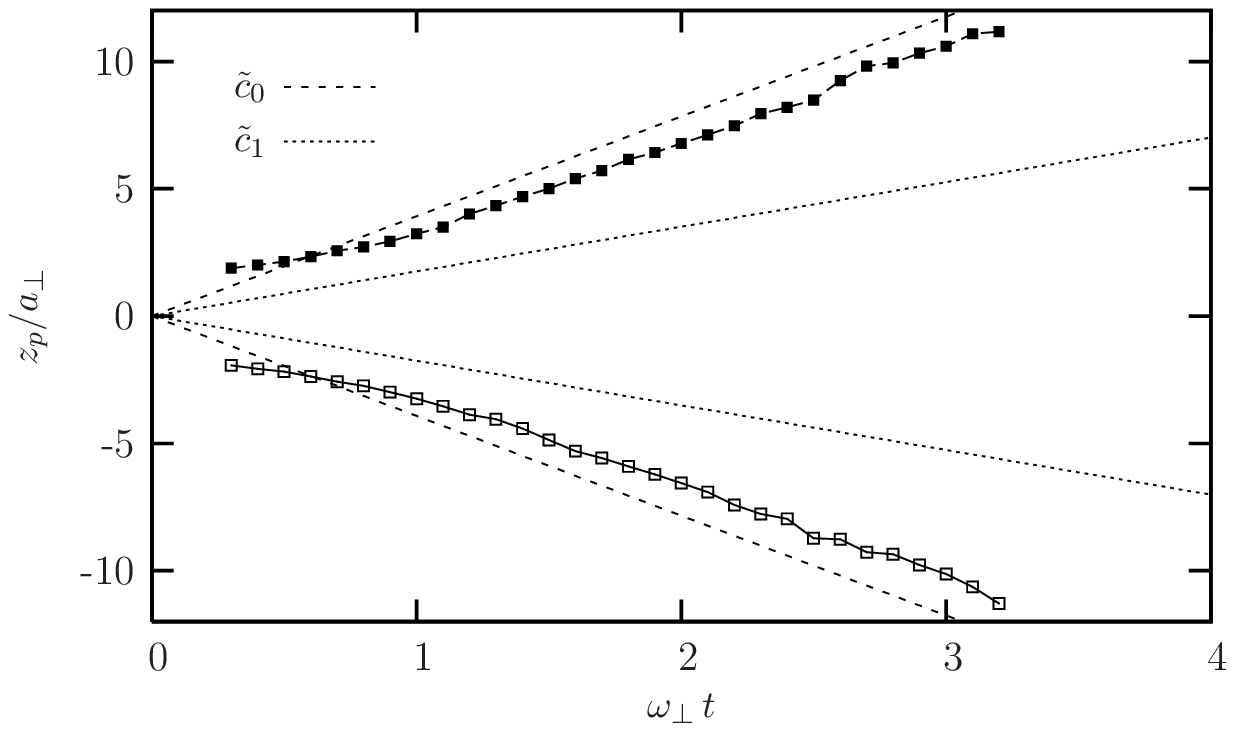}
\includegraphics[width=0.7\columnwidth,clip=true]{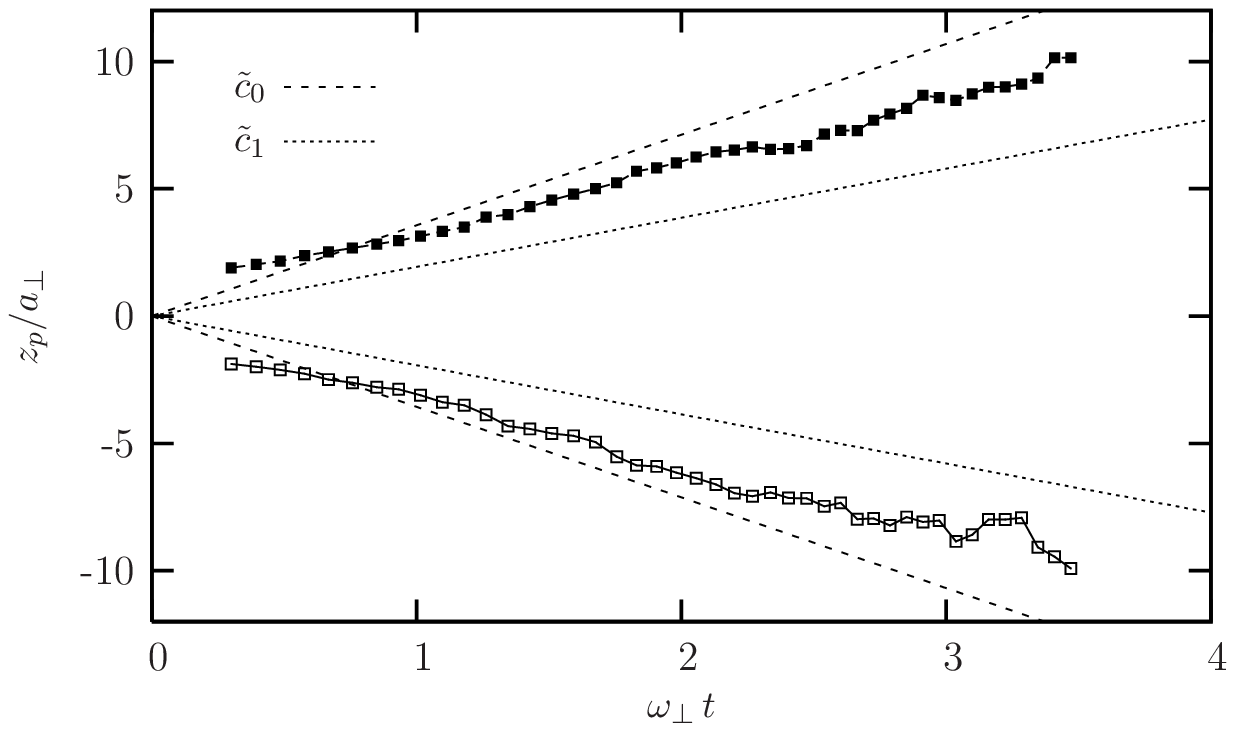}
\end{center}
\caption{\label{fig:peaks} Location $z_p/a_{\perp}$ of the density peaks 
in $\tilde{n}(z)$ 
as a function of time (in units of $1/\omega_{\perp}$). The top, middle,
and bottom panels correspond to $a=-1.5\times 10^4$, 80, and
$2\times10^4$ Bohr radii. The dashed and dotted lines refer to 
propagation at the zero-sound velocity $\tilde c_0$ and at the first-sound 
velocity $\tilde{c}_1$, respectively.}
\end{figure}

The propagation of a density perturbation is evaluated by performing
the following numerical experiment. First we construct the equilibrium
density profiles according to  equations (\ref{eq:nsigma})-(\ref{eq:Ueff}) 
for a given confinement configuration and interaction
strength. A Gaussian potential
$U_0\exp(-z^2/2w^2)$ with amplitude $U_0=10\hbar\omega_{\perp}$ and
width $w=(\hbar/(m\omega_{\perp}))^{1/2}$ is then switched on at
the centre of the trap to simulate a laser beam expelling particles  
from the centre of the cloud.
At later times these density distortions are seen to
move towards the ends of the trap.
Since each perturbation has velocity mainly in the $z$ direction, we
focus on the evolution of the integrated density profile
$\tilde{n}(z)=\int n(\bi{r})\,d^2r_{\perp}$.  Experimentally, this
density can be obtained  directly from \textit{in-situ} measurements
of the three-dimensional density profiles.  In
figure \ref{fig:prof_shifted} we plot  $\tilde{n}(z)$  at several times
for  a trap with $\lambda=0.1$ and for three values of the scattering
length, namely $a=-1.5\times 10^4$, 80, and $2\times 10^4$ Bohr radii 
$a_0$. At
larger attractions the gas collapses, whereas for
larger repulsions the effects of demixing become noticeable.
We follow the positions of the main density peaks created by the 
Gaussian potential
and plot in figure \ref{fig:peaks} their evolution in time. After
a hole in the density has been created, we observe that density
pulses travel with constant speed in the neighbourhood of the trap
centre. For $a=-1.5\times 10^4\,a_0$ and $2\times 10^4\,a_0$ 
it can be seen that
the slope of the peak positions clearly departs from
$\tilde{c}_{0}$ and moves closer to $\tilde{c}_1$, as the
hydrodynamic regime is being approached.


\begin{figure}
\includegraphics[width=\columnwidth,clip=true]{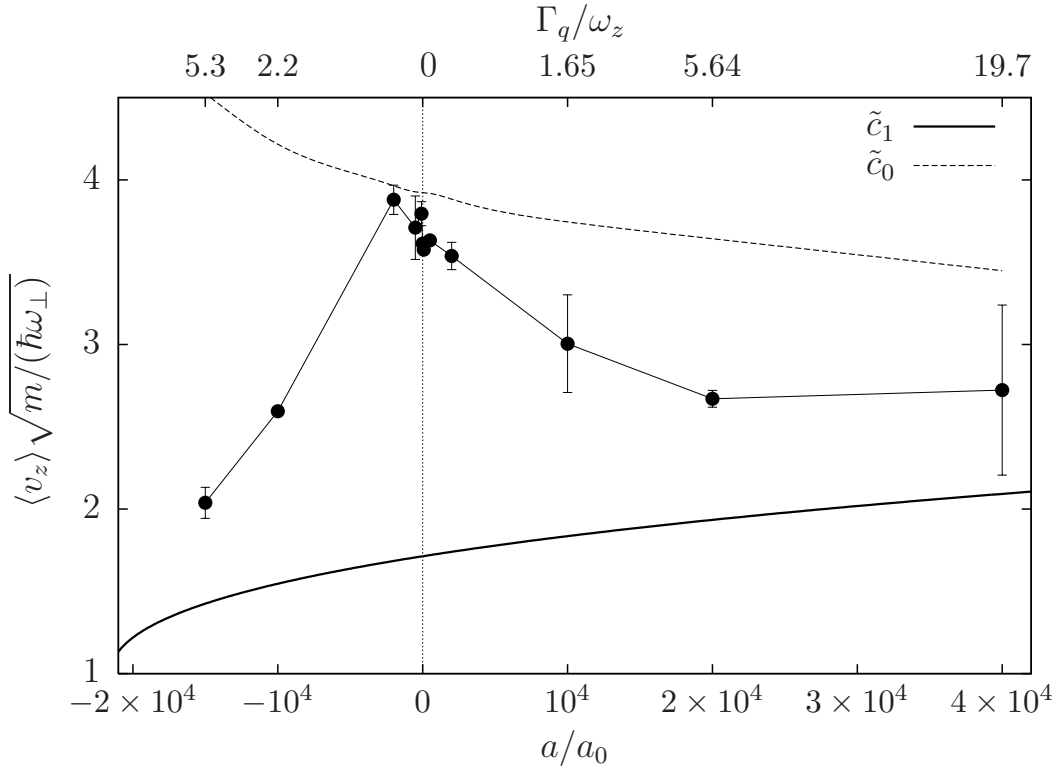}
\caption{\label{fig:soundone_1}Average velocity $\langle v_z\rangle$
(in units of $\sqrt{\hbar\omega_{\perp}/m}$) as a function of the
scattering length $a$  (in units of the Bohr radius $a_0$) for
$\lambda=0.1$. The thick solid line and the dashed line show the sound 
velocity calculated in the absence of damping as for first sound and
zero sound, respectively. The dots with error bars are the results of
the VLE simulation, the errors being estimated from the noise in the slope 
of the density peak positions in figure \ref{fig:peaks}. The thin solid
line is just a guide to the eye. The scale at the top gives the
quantum collision rate $\Gamma_q$ (in units of $\omega_z$).} 
\end{figure}

The results for the sound velocity as a function of the coupling
are summarized in
figure \ref{fig:soundone_1}, where we show the mean value $\langle v_z
\rangle = \frac{1}{2}(\langle v_z^{(s)}\rangle + \langle
v_z^{(\bar{s})} \rangle)$ of the time-averaged velocities of the  two
components of the mixture as obtained from the solution of the VLE.  
These have been evaluated over a time interval  $\Delta
t = 1.5/\omega_{\perp}$ starting at $t_0=1/\omega_{\perp}$. 
In the same figure we also report the results of
the analytical models introduced
in Sec.~\ref{sec:sound_propa}, \textit{i.e.}
the zero-sound velocity $\tilde c_0$ in a strongly elongated trap (cf.
equation  (\ref{eq:zerosound})), and the first-sound velocity $\tilde
c_1$ in a strongly elongated trap as given in equation (\ref{eq:cstrap})
with $n^{(s)}=n^{(\bar{s})}$. In the latter case we have used the
zero-temperature equation of state (\ref{eq:eos}) and set $\mu$ to the
value obtained in the VLE simulation. 
By solving the VLE in the absence of collisions we also
find that the speed of the density
perturbations is well defined for both positive and negative couplings
and follows the trend of $\tilde{c}_0$ as a function of $a$ (not
shown in the figure). 
We emphasis that in the absence of collisions zero sound is a well
defined collective excitation in the gas under quasi-1D confinement
even when the interactions between the particles are attractive, contrary to the case of a homogeneous 3D gas where these excitations are stable
only for interparticle repulsions \cite{Pines94}. On the other hand, in
the case of repulsive interactions our results for $\tilde c_0$ are very 
close to those obtained from equation (\ref{eq:eta3D}) for the homogeneous gas.

From the full solution of the VLE including collisions, we observe
that at low coupling the velocity of the density perturbations is
close to $\tilde c_0$ within the error bars.  In this regime the value
of the sound velocity is dominated by the mean-field interactions and
their effect on sound propagation can be understood as mainly
resulting from changes in the local density: in particular, for
attractive coupling the Fermi velocity and the sound velocity rise as
the density of the gas is increased.  At stronger coupling strength
the sound velocity drops for both positive and negative coupling, as
opposed to the behaviour of the collisionless velocity $\tilde{c}_0$.
This drop is more pronounced for attractive coupling as the frequency
of collisions increases with $|a|$ as well as with the gas density,
whereas at strongly repulsive interactions the collisions increase
more slowly as the two components approach spatial demixing.  This
effect is not included in the calculation of $\tilde{c}_1$.  At
strongly attractive interactions the frequency of collisions is much
larger than $\omega_z$ and therefore the gas is strongly collisional.
The difference between $\langle v_z\rangle$ and $\tilde{c}_1$ at
$a=-1.5\times 10^4\,a_0$ may be attributed to the appreciable
amplitude of the density perturbations as well as to the confinement
in the axial direction. At strong coupling, although the rate of
collisions in the axially confined inhomogeneous cloud varies at each
point in space during the propagation of the pulse, the sound wave
travels without observable damping in the neighbourhood of the trap
centre. 

\begin{figure}
\includegraphics[width=0.9\columnwidth,clip=true]{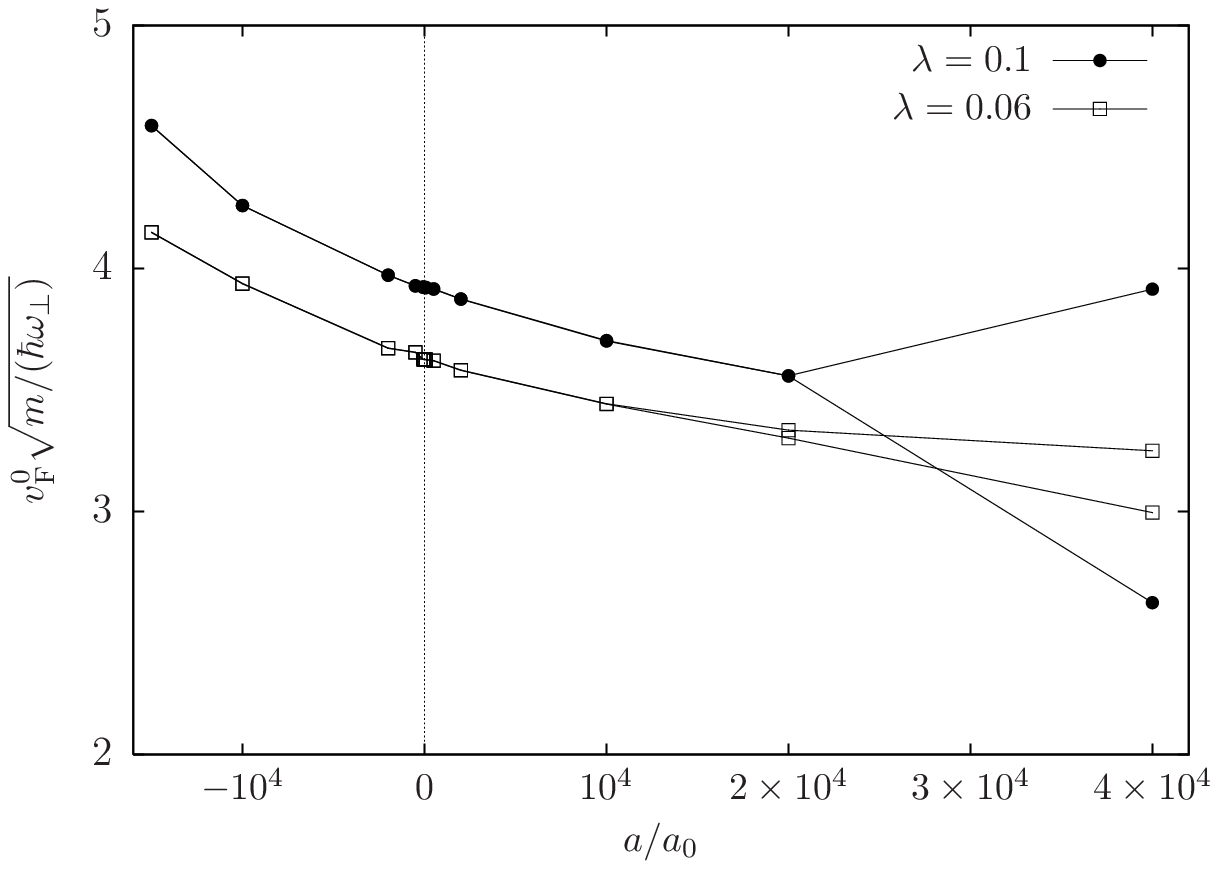}\\
\includegraphics[width=0.9\columnwidth,clip=true]{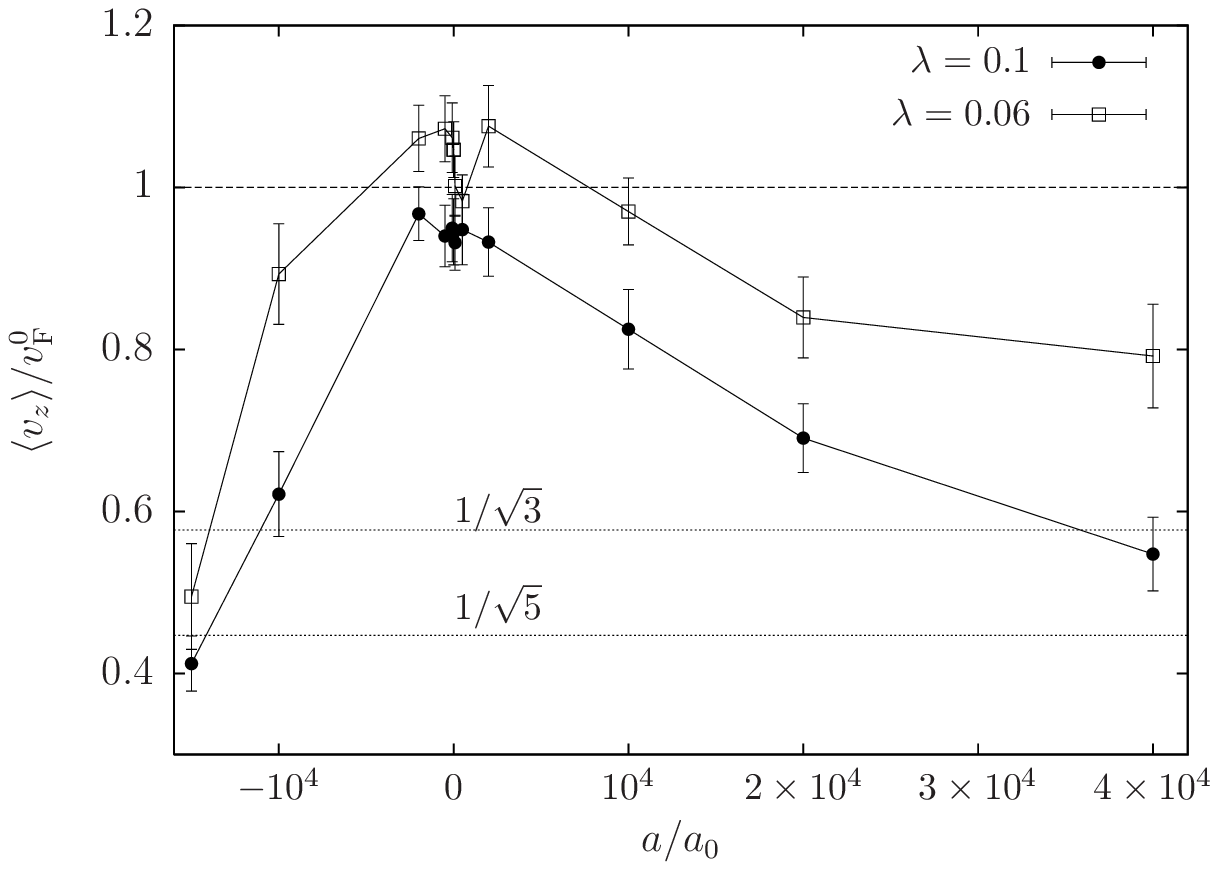}
\caption{\label{fig:vmedia}Sound velocity in elongated traps with  
$\lambda= 0.1$ or  $\lambda=0.06$. Top panel: Local Fermi velocity
$v_{\mathrm{F}}^0$ at the cloud centre (in units of
$\sqrt{\hbar\omega_{\perp}/m}$) as a function of $a$ (in units of the
Bohr radius). Bottom panel: Average sound velocity (in units of
$v_{\mathrm{F}}^0$) as a function of $a$ (in units of the Bohr
radius).}
\end{figure}

The behaviour of $\langle v_z\rangle$ for two values of the anisotropy
parameter at
fixed $\omega_{\perp}$ is illustrated in figure \ref{fig:vmedia}.  In
the top panel we report the Fermi velocity at the cloud centre 
for both components, showing that in the more elongated
cloud the density of the
gas and thus the Fermi velocity are lower.  Therefore, at a fixed
value of $a$ the collisionality of the gas with $\lambda=0.06$  is
lower  and  $\langle v_z\rangle/v_{\mathrm{F}}^0$ is higher (bottom
panel of figure \ref{fig:vmedia}). On the other hand, if we fix the
average trap frequency and thus the density at the centre, a
deformation of the trap would increase the number of collisions
occurring during sound propagation and therefore facilitate the
transition towards the hydrodynamic limit \cite{Capuzzi2005a}.
The effect of demixing at strong repulsive interaction is also
illustrated in the top panel of figure \ref{fig:vmedia}.

\section{\label{sec:summary}Summary and concluding remarks}

We have studied the propagation of sound waves in a mixture of
fermionic atoms confined inside a strongly elongated cigar-shaped trap
and exposed to the role of collisions in determining the speed of
propagation. We have compared our results from the numerical
solution of the Vlasov-Landau equations with
two limiting behaviours: zero sound and first sound propagating in
an infinitely long cylindrical cloud with a very tight transverse
confinement. We have found substantial agreement between
the velocity of density perturbations in the collisionless regime and
in the highly collisional regime with those of zero sound and first
sound, respectively. The differences may be attributed to the finite amplitude of
the density perturbations that we have produced and to the axial confinement. 

It would be interesting to study the effect of the presence of a
superfluid component on the zero-to-first sound transition at finite
temperature. This can be done by using suitable transport
equations for a trapped Fermi gas below the superfluid transition
temperature, as already treated by Urban and Schuck \cite{Urban2005a}.
 
\ack
This work has been partially supported by an Advanced Research Initiative 
of Scuola Normale Superiore di Pisa and by the Istituto Nazionale di 
Fisica della Materia within the Initiative ``Calcolo Parallelo''.

\appendix
\section{Zero sound in a cylindrical trap}
We apply the transport equation   
(\ref{eq:vlasov}) with $C^{(s)}=0$ and
$V_{\rm{ext}}=m\omega_{\perp}^2 r_{\perp}^2/2\equiv\tilde{V}$ 
to evaluate the propagation of zero sound in a tight cylindrical trap. 
A density perturbation 
travels mainly along the axis of the cylinder when the radial
confinement is strong, and 
the distribution function can be written as
\begin{equation}
f^{(s)}({\bi{r}},{\bi{p}})=f_0^{(s)}({\bi{r}}_\perp,{\bi{p}})+
f_1^{(s)}({\bi{r}}_\perp,{\bi{p}})
e^{i( k z-\omega t)}
\end{equation}
where $f_0^{(s)}=\theta(\mu^{(s)}-\frac{p^2}{2m}-
\tilde V(r_\perp)-gn_0^{(\bar s)}(r_\perp))$ is the equilibrium distribution.
Around the centre of the azimuthal plane one can write
$
f_0^{(s)}=\theta(\tilde\mu^{(s)}-\frac{p_z^2}{2m}
-\frac{p_\perp^2}{2m}-\frac{1}{2}m\tilde\omega_\perp^2r_\perp^2)
$
with $\tilde\mu^{(s)}=\mu^{(s)}-gn_0^{(\bar s)}(0)$
and $\tilde\omega_\perp=\omega_\perp (1-3gn_0^{(\bar s)}(0)/(2\mu))^{1/2}$.
Only the first level in the
radial harmonic potential with frequency $\tilde\omega_\perp$
is occupied by the fermions in a very tight radial confinement, and the distribution function takes
the simple form $f_0^{(s)}=\theta(\tilde\mu^{(s)}-\frac{p_z^2}{2m}-
\hbar\tilde\omega_\perp)$.

The terms in the transport equation involving
the derivatives with respect to $r_\perp$ and $p_\perp$ disappear if
$f_1^{(s)}$ is a function of the total energy analogously to
$f_0^{(s)}$. Equation (\ref{eq:vlasov}) then becomes
\begin{equation}
\left(\frac{k\, p_z}{m}-\omega\right)f_1^{(s)}-
\frac{k\, p_z}{m}
\frac{\partial f_0^{(s)}}{\partial\varepsilon}
\int \frac{d{\bi{p}}}
{(2\pi\hbar)^3} g f_1^{(\bar s)}=0\,,
\label{sistema2}
\end{equation}
where we have set $\varepsilon=p_z^2/(2m)$.
The form of equation (\ref{sistema2}) suggests 
to write $f_1^{(s)}$ as
proportional to
$\nu\,\partial f_0^{(s)}/\partial \varepsilon $,
$\nu$ being defined as 
\begin{equation}
\nu=\cases{\frac{1}{1-\eta}& if $k p_z > 0$\\
\frac{1}{1+\eta}& if $k p_z < 0$}
\end{equation}
with $\eta=\omega/(k v_{\mathrm{F}}^{(s)})$ and $v_{\mathrm{F}}^{(s)}=
\sqrt{2(\tilde \mu^{(s)}-\hbar\tilde\omega_\perp)/m}$.
Thus equation (\ref{sistema2}) can be rewritten as 
\begin{equation}
1+\frac{g}{1-\eta^2}\int 
\frac{d^3p}{(2\pi\hbar)^3} \delta(\tilde\mu^{(\bar s)}-\frac{p_z^2}{2m}-
\hbar\tilde\omega_\perp)=0\,.
\label{quasi}
\end{equation}
Finally, we obtain
\begin{equation}
\eta=\sqrt{1+\frac{2\hbar a}{\pi\tilde{a}_\perp^2 m v_{\mathrm{F}}^{(\bar s)}}}\,.
\end{equation}
by integrating over all $p_z$ and
$p_\perp\in[0,\sqrt{2\hbar m\tilde\omega_\perp}]$. This is equation 
(\ref{eq:zerosound}) in the main text.


\section*{References}

\begin{thebibliography}{xx}

\bibitem{Gensemer2001a}
Gensemer S~D and Jin D~S  2001 {\em Phys. Rev. Lett.} {\bf 87}~173201

\bibitem{Kinast2004a}
Kinast J, Hemmer S~L, Gehm M~E, Turlapov A and Thomas J~E  2004 {\em Phys. Rev.
  Lett.} {\bf 92}~150402

\bibitem{Bartenstein2004a}
Bartenstein M, Altmeyer A, Riedl S, Jochim S, Chin C, Denschlag J~H and Grimm R
   2004 {\em Phys. Rev. Lett.} {\bf 92}~203201

\bibitem{DeMarco2002a}
DeMarco B and Jin D~S  2002 {\em Phys. Rev. Lett.} {\bf 88}~040405

\bibitem{Loftus2002a}
Loftus T, Regal C~A, Ticknor C, Bohn J~L and Jin D~S  2002 {\em Phys. Rev.
  Lett.} {\bf 88}~173201

\bibitem{Regal2003b}
Regal C~A and Jin D~S  2003 {\em Phys. Rev. Lett.} {\bf 90}~230404

\bibitem{Toschi2003a}
Toschi F, Vignolo P, Succi S and Tosi M~P  2003 {\em Phys. Rev. A} {\bf
  67}~041605

\bibitem{Toschi2004a}
Toschi F, Capuzzi P, Succi S, Vignolo P and Tosi M~P  2004 {\em J. Phys. B}
  {\bf 37}~S91

\bibitem{Ho2004a}
Ho T~L  2004 {\em Phys. Rev. Lett.} {\bf 92}~090402

\bibitem{Heiselberg2005b}
Heiselberg H  2005.
\newblock \textit{Preprint} cond-mat/0503101

\bibitem{Capuzzi2006a}
Capuzzi P, Vignolo P, Federici F and Tosi M~P  2005.
\newblock \textit{Preprint} cond-mat/0509323

\bibitem{Urban2005a}
Urban M and Schuck P  2005.
\newblock \textit{Preprint} cond-mat/0509373

\bibitem{Heiselberg2005a}
Heiselberg H  2004.
\newblock \textit{Preprint} cond-mat/0409077

\bibitem{Pines94} See \textit{e.g.} 
Pines D and Nozi{\'e}res P  1994 {\em The Theory of Quantum Liquids} Vol.~1
  (New York: Addison-Wesley)

\bibitem{Abel1966a}
Abel W~R, Anderson A~C and Wheatley J~C  1966 {\em Phys. Rev. Lett.} {\bf
  17}~74

\bibitem{Andrews1997a}
Andrews M~R, Kurn D~M, Miesner H~J, Durfee D~S, Townsend C~G, Inouye S and
  Ketterle W  1997 {\em Phys. Rev. Lett.} {\bf 79}~553

\bibitem{Akdeniz2003b}
Akdeniz Z, Vignolo P and Tosi M~P  2003 {\em Phys. Lett. A} {\bf 311}~246

\bibitem{Apostol1992a}
Apostol M and Malomed B~A  1992 {\em Phys. Rev. B} {\bf 45}~4509

\bibitem{Hernandez2002a}
Hern\'andez E~S  2002 {\em J. Low Temp. Phys.} {\bf 127}~153

\bibitem{Salasnich2002a}
Salasnich L, Parola A and Reatto L  2002 {\em Phys. Rev. A} {\bf 65}~043614

\bibitem{Capuzzi2001a}
Capuzzi P and Hern{\'a}ndez E~S  2001 {\em Phys. Rev. A} {\bf 63}~063606

\bibitem{Bruun2001b}
Bruun G~M  2001 {\em Phys. Rev. A} {\bf 63}~043408

\bibitem{Vichi1999a}
Vichi L and Stringari S  1999 {\em Phys. Rev. A} {\bf 60}~4734

\bibitem{Bruun1999b}
Bruun G~M and Clark C~W  1999 {\em Phys. Rev. Lett.} {\bf 83}~5415

\bibitem{Amoruso2000a}
Amoruso M, Meccoli I, Minguzzi A and Tosi M~P  2000 {\em Eur. Phys. J. D} {\bf
  8}~361

\bibitem{Chin2004a}
Chin C, Bartenstein M, Altmeyer A, Riedl S, Jochim S, {Hecker Denschlag} J and
  Grimm R  2004 {\em Science} {\bf 305}~1128

\bibitem{Zwierlein2005a}
Zwierlein M~W, Abo-Shaeer J~R, Schirotzek A, Schunck C~H and Ketterle W  2005
  {\em Nature} {\bf 435}~1047

\bibitem{Greiner2003a}
Greiner M, Regal C~A and Jin D~S  2003 {\em Nature} {\bf 426}~537

\bibitem{Jochim2003b}
Jochim S, Bartenstein M, Altmeyer A, Hendl G, Riedl S, Chin C, Denschlag J~H
  and Grimm R  2003 {\em Science} {\bf 302}~2101

\bibitem{Zwierlein2003a}
Zwierlein M~W, Stan C~A, Schunck C~H, Raupach S~M~F, Gupta S, Hadzibabic Z and
  Ketterle W  2003 {\em Phys. Rev. Lett.} {\bf 91}~250401

\bibitem{Capuzzi2005a}
Capuzzi P, Vignolo P and Tosi M~P  2005 {\em Phys. Rev. A} {\bf 72}~013618

\end{thebibliography}
\end{document}